\documentstyle[prd,aps]{revtex}

\makeatletter

\providecommand{\LyX}{L\kern-.1667em\lower.25em\hbox{Y}\kern-.125emX\@}

\makeatother

\begin{document}

\title{Conservative general relativistic radiation 
hydrodynamics in spherical symmetry and comoving coordinates}

\author{
Matthias Liebend\"{o}rfer$^{1,2,3}$, Anthony Mezzacappa$^{2}$,
and Friedrich-Karl Thielemann$^{1,2}$
}

\address{
$^{1}$ Department of Physics and Astronomy, University of Basel,
4056 Basel, Switzerland \\
$^{2}$ Physics Division, Oak Ridge National Laboratory, Oak Ridge,
Tennessee 37831-6354 \\
$^{3}$ Department of Physics and Astronomy, University of Tennessee,
Knoxville, Tennessee 37996-1200
}

\maketitle
\begin{abstract}
The description of general relativistic radiation hydrodynamics in
spherical symmetry is presented in natural coordinate choices.
For hydrodynamics, comoving coordinates are chosen, and the
momentum phase space for the radiation particles is described
in comoving frame four-momenta. We also investigate a description
of the momentum phase space in terms of the particle impact parameter
and energy at infinity and derive a simple approximation to
the general relativistic Boltzmann equation. Further developed
are, however, the exact equations in comoving coordinates, because
the description of
the interaction between matter and radiation particles is best
described in the closely related orthonormal basis comoving with
the fluid elements. We achieve a conservative and concise
formulation of radiation hydrodynamics that is well suited for
numerical implementation by a variety of methods. The contribution
of radiation to the general relativistic jump conditions at shock
fronts is discussed, and artificial viscosity is consistently
included in the derivations in order to support approaches relying
on this option.
\end{abstract}


\section{Introduction }

Astrophysical
knowledge is gained by the observation of luminous objects and the 
search for scenarios that explain the 
observations based on accepted physics. Often,
numerical simulations are required to determine if a detailed
scenario really produces the given observation. In supernova theory,
for example,
the transition from analytical considerations of core collapse and 
explosion
(Colgate and Johnson \cite{Colgate_Johnson_60}) to numerical 
simulations (Colgate
and White \cite{Colgate_White_66}) was made very early, when 
computers became available. General relativistic 
simulations of core collapse
followed immediately (May and White \cite{May_White_67}). These 
simulations
were based on the newly derived Einstein equations in spherical 
symmetry (Misner
and Sharp \cite{Misner_Sharp_64}) that served as a basis for many 
later investigations,
including also the present paper. In comoving coordinates, 
general
relativistic radiation transport was first formulated by Lindquist 
\cite{Lindquist_66}.
Here, we reformulate general relativistic radiation hydrodynamics 
in light of modern numerical algorithms requiring
conservation laws, and hope to reveal some of the beauty in the 
spherically symmetric case that may have remained hidden.

However, comoving coordinates are only one possible choice out
of a variety of 3+1 decompositions enabled by the covariance of general
relativity in four-dimensional space-time (Arnowitt, Deser, and 
Misner \cite{Arnowitt_Deser_Misner_62}; Smarr and York \cite{Smarr_York_78}).
The Boltzmann transport
equation is usually split into a left-hand side and a right-hand side.
The left-hand side is the directional derivative of the distribution
function along trajectories of free particle propagation. This
derivative is equated to the changes in the distribution
function due to collisions. Thus, the right-hand side accounts
for changes in the radiation particle distribution function
owing to particles that are created, annihilated, or scattered
into new states in the momentum phase space.

After the choices of a 3+1 decomposition and a basis
in the momentum phase space have been made, the directional
derivative along the phase flow can be expressed in terms of
partial derivatives of the
distribution function with respect to the coordinates.
The complexity of the left-hand side as well as the
complexity of the collision term depends on the coordinate choice.
A general discussion of radiation transport in spherically symmetric 3+1
decomposition has been provided by Mezzacappa and Matzner
\cite{Mezzacappa_Matzner_89}. Maximal slicing was chosen
and the particle distribution function was
described by the four-momenta measured in the frame of an observer
comoving with the matter. This choice eases the evaluation
of angle dependent cross sections in the collision term 
(Mihalas \cite{Mihalas_86},
Mezzacappa and Bruenn \cite{Mezzacappa_Bruenn_93b}).
Schinder and Bludman 
\cite{Schinder_Bludman_89} chose polar slicing in the 3+1
decomposition and a tangent ray approach in the description
of the momentum phase space. This choice avoids partial
derivatives with respect to the momentum space variables
on the left-hand side of the Boltzmann equation.

In this work we explore the most natural 3+1 decomposition,
one that is not enforced by an artificial external slicing condition. The
time slices in the orthogonal comoving coordinates used by Misner and
Sharp \cite{Misner_Sharp_64} are attached to the dynamical motion of the matter.
As for the basis of the momentum space we first investigate the
description in the comoving frame. In another part, we
also develop a formulation in the spirit of the tangent ray
approach and draw the connections between these two choices.
But first of all, we continue with a motivation of the
comoving coordinate choice.

The most obvious advantage is already contained in the word 
{\em comoving}. A given coordinate interval
moves with the matter and adjusts
to the location where the matter is. Comoving coordinates
ease the description of the internal physical state of 
fluid elements because numerically difficult advection terms do
not enter the hydrodynamics equations as in other coordinate
systems. Gravitational collapse in spherical symmetry is 
a natural application of comoving coordinates.
With regard to radiation transport, the comoving frame, which
is preferred for the evaluation of the integrals over the
angle dependent emission, absorption, and scattering kernels,
is collinear with the comoving coordinates such that Lorentz
transformations are avoided in the collision term.

On the other hand, the orthogonal comoving coordinates show an 
important drawback
in general relativity. After the formation of a black hole,
observers may fall within finite time into the physical singularity 
at the
symmetry center. In addition, a coordinate singularity forms at 
the Schwarzschild
horizon. Simulations of the outer regions cannot proceed because of 
these singularities
in the computational domain.

To circumvent this difficulty, interest focused on singularity 
avoiding time
slicing. For example, simulations of Wilson \cite{Wilson_79} and
Shapiro and Teukolsky \cite{Shapiro_Teukolsky_80}
were carried out in maximal slicing. The collapse to a black hole can be followed 
beyond
the appearence of trapped surfaces in these coordinates. At late times 
however,
the grid is ``sucked down'' the black hole, leading to unsatisfactory 
resolution
in the physically interesting region outside the event horizon. 
Polar slicing
shows even stronger singularity avoidance (Bardeen and Piran 
\cite{Bardeen_Piran_83}).
The computational domain stays outside the apparent horizon for 
all times.
This time slicing was implemented by Romero et al. 
\cite{Romero_et_al_96} in
orthogonal coordinates. Schinder et al. 
\cite{Schinder_Bludman_Piran_88} found
an interesting compromise by introducing nonzero shift vectors to get
comoving coordinates in polar slicing. Orthogonality of the 
coordinates, however,
had to be abandoned. The idea of using observer time coordinates 
suggested by Hernandez
and Misner \cite{Hernandez_Misner_66} has been used by 
Miller and Motta \cite{Miller_Motta_89} and Baumgarte et 
al. \cite{Baumgarte_Shapiro_Teukolsky_95}
in a singularity avoiding code. Although coordinate and physical
singularities are not encountered
in singularity avoiding schemes, the coefficients of the metric can 
increase
without limits for late times (Petrich et al. 
\cite{Petrich_Shapiro_Teukolsky_86}).
Alternatively, the computational domain can also be limited to the 
region outside
of the event horizon by the choice of appropriate shift vectors 
(Liebend\"orfer
and Thielemann \cite{Liebendoerfer_Thielemann_98}). A free choice of 
time
slicing is then reestablished even in the presence of singularities.

One argument against the use of comoving orthogonal coordinates is
the difficulty to extend
them to multiple spatial dimensions because of grid entangling.
However, one-dimensional simulations offer the opportunity to include 
general
relativity, exact Boltzmann radiation transport, and sophisticated 
physics for
emission, absorption, and scattering at once. While the individual 
pieces are
well represented in the literature, dynamical simulations with all 
ingredients
are very difficult and about to become current state of the art.

Many spherically symmetric simulations of compact objects were based 
on the
comoving orthogonal coordinates of Misner and Sharp 
\cite{Misner_Sharp_64}.
Finite difference schemes were constructed by May and White 
\cite{May_White_67},
Van Riper \cite{VanRiper_79}, Bruenn \cite{Bruenn_85},
Rezzolla and Miller \cite{Rezzolla_Miller_94}, and Swesty
\cite{Swesty_95}. An approximate Riemann
solver was constructed by Yamada \cite{Yamada_97}. However, the 
appearance
of the time dependent metric in the general relativistic equations 
prevented
an immediate application of numerical methods developed for 
conservatively
formulated hydrodynamics. A formulation of the dynamics in terms of 
conservation
laws has several benefits (\emph{i}) Fundamental conservation laws 
are
also valid at discontinuities and allow an accurate numerical 
solution - e.g.,
for the propagation of shock waves. This is not the case with 
arbitrary finite
difference approximations. (\emph{ii}) The integration of a conserved 
quantity
over adjacent fluid elements does not depend on the fluxes at the 
enclosed boundaries.
Thus, discretization errors have mainly local influence. This 
advantage is important
for the implicit solution of problems involving scale differences of 
many orders
of magnitude. (\emph{iii}) The integration over the whole 
computational domain
of the single-fluid-element conservation laws leads to the total 
conservation
of fundamental physical quantities, independent of the resolution 
in
the conservative finite difference representation. (\emph{iv}) An 
overwhelming
variety of generalized investigations and numerical methods for the 
solution
of hyperbolic conservation laws already exists (see, e.g., Davis 
\cite{Davis_88} and
references therein). In analogy to the work of Romero et al. 
\cite{Romero_et_al_96}
in polar slicing, Liebend\"orfer and Thielemann 
\cite{Liebendoerfer_Thielemann_98}
formulated conservative equations for hydrodynamics in the comoving 
frame.
This approach is extended here to include radiation 
transport. The description (with the omission of neutrino
back reaction to the fluid metric) has successfully
been used in a general relativistic simulation of stellar core 
collapse, bounce and postbounce evolution, based on multigroup Boltzmann
neutrino transport (Liebend\"orfer et al.
\cite{Liebendoerfer_00,Liebendoerfer_et_al_00}).


\section{Conservative Einstein 
Equations\label{subsection_conservative_formulation}}

The left-hand side of the Einstein field equation, the Einstein 
tensor,
is determined by the metric. Following Misner and Sharp 
\cite{Misner_Sharp_64},
we select in spherical symmetry
\begin{equation}
ds^{2}=-\alpha ^{2}dt^{2}+\left( \frac{r'}{\Gamma }\right) ^{2}da^{2}
+r^{2}\left( d\vartheta ^{2}+\sin ^{2}\vartheta d\varphi ^{2}\right) 
, \label{eq_comoving_metric}
\end{equation}
where \( r \) is the areal radius and \( a \) is a label corresponding
to an enclosed rest mass
(the prime
denotes a derivative with respect to \( a \): \( r'=\partial r/\partial 
a \)).
The proper time lapse of an observer attached to the motion of rest 
mass is
related to the coordinate time \( dt \) by the lapse function \( 
\alpha  \).
The angles \( \vartheta  \) and \( \varphi  \) describe a two-sphere.

The right-hand side of the Einstein equations is given by the fluid- 
and
radiation stress-energy tensor, \( T \), which, in a comoving
orthonormal basis, has the components
\begin{eqnarray}
T^{tt} & = & \rho \left( 1+e\right) \nonumber \\
T^{ta}=T^{at} & = & q \nonumber \\
T^{aa}=T^{\vartheta \vartheta }=T^{\varphi \varphi } & = & p.
\label{eq_stress_energy_tensor}
\end{eqnarray}
The total energy is expressed in terms of the rest mass density, \( 
\rho  \),
and the specific
energy, \( e \). The isotropic pressure is denoted by \( p \), and 
radial net
energy transport is included by the nondiagonal component \( q \). 
Note our
convention that energy density and pressure contain the contribution 
from
the radiation field as well as the matter.

In the Appendix, we rederive a system of equations
equivalent to the Einstein equations in spherical symmetry, closely
following the guideline of Misner et al. 
\cite{Misner_Thorne_Wheeler_73}.
The only difference to previous work is the omission of any 
approximations
in the derivation. Moreover, the Appendix gives many useful relations
for what follows. The final system of exact equations
reads
\begin{eqnarray}
\frac{\partial r}{\alpha \partial t} & = & u\label{misner_sharp_drdt} \\
\frac{\partial m}{\alpha \partial t} & = & -4\pi r^{2}\left( up+\Gamma q\right) 
\label{misner_sharp_dmdt} \\
\frac{\partial u}{\alpha \partial t} & = & \frac{\Gamma ^{2}}{\alpha }\frac{\partial 
\alpha }{\partial r}-\frac{m+4\pi 
r^{3}p}{r^{2}}\label{misner_sharp_dudt} \\
r' & = & \frac{\Gamma }{4\pi r^{2}\rho }\label{misner_sharp_r'} \\
m' & = & \Gamma \left( 1+e\right) +u\frac{q}{\rho 
}\label{misner_sharp_m'} \\
\alpha ' & = & -\frac{\alpha }{\left( 1+e+p/\rho \right) }\left[ 
\frac{p'}{\rho }+\frac{\partial}{\alpha \partial t}\left( \frac{1}{4\pi r^{2}\rho 
}\frac{q}{\rho }\right) \right] .\label{misner_sharp_a'} 
\end{eqnarray}
The velocity \( u \) 
is equivalent to the \( r \) component of the fluid four-velocity 
as observed
from a frame at constant areal radius (May and White 
\cite{May_White_67}).
In the special relativistic limit, \( \Gamma =\sqrt{1+u^{2}-2m/r} \)
becomes the Lorentz factor corresponding to the boost between
the inertial and the comoving observers.
The gravitational mass \( m \) is the total energy enclosed 
in the
sphere at rest mass \( a \). Its change is determined by the surface 
work, involving
pressure \( p \) and velocity \( u \), and the boundary luminosity \( 
L=4\pi r^{2}q \). 

We will now show
that Eqs. (\ref{misner_sharp_drdt})-(\ref{misner_sharp_a'}) can be 
written in conservative form. Optimal for 
a numerical implementation is a formulation that contains quantities
that are either conserved or local. Since the rest mass as
independent variable is trivially conserved, we look out for the
conservation of volume, energy and momentum. Equation
(\ref{misner_sharp_r'}) leads to the definition of the integral
\begin{equation}
V\equiv\frac{4\pi }{3}r^{3} = \int _{0}^{a}\frac{\Gamma }{\rho 
}da \label{eq_conserved_volume}
\end{equation}
as volume and Eq. (\ref{misner_sharp_m'}) suggests the definition
of the integral
\begin{equation}
m = \int _{0}^{a}\left[ \Gamma \left( 1+e\right) +\frac{uq}{\rho 
}\right] da\label{eq_conserved_energy} 
\end{equation}
as gravitational mass, equivalent to total enclosed energy.
For the radial momentum we choose \( u\left( 
1+e\right) +\Gamma q/ \rho \) and show later that it indeed leads
to a conservation equation. Borrowing 
the notation from Romero et al. \cite{Romero_et_al_96}, we rename the
candidates for conserved quantities and define
specific volume, specific energy and specific momentum as
\begin{eqnarray}
\frac{1}{D} & = & \frac{\Gamma }{\rho }\label{eq_specific_volume}\\
\tau & = & \Gamma e+\frac{2}{\Gamma +1}\left( 
\frac{1}{2}u^{2}-\frac{m}{r}\right) +\frac{uq}{\rho }
\label{equatino_specific_energy}\\
S & = & u\left( 1+e\right) +\Gamma \frac{q}{\rho }.
\label{eq_specific_momentum}
\end{eqnarray}
In the nonrelativistic limit, we retrieve with \( \alpha =\Gamma =1 \)
the familiar specific volume \( D = 1/\rho \), the sum of the specific
internal, kinetic, and gravitational energy \( \tau = e + u^2/2 - m/r \), and
the specific radial momentum \( S = u \). With the help of Eqs. 
(\ref{misner_sharp_drdt}) and (\ref{misner_sharp_r'}), the time
derivative of \( 1/D  \) leads to the continuity equation (\ref{eq_continuity}).
An energy equation is obtained by taking the time derivative of specific
total energy \( 1+\tau \) and substituting
Eqs. (\ref{misner_sharp_dmdt}) and (\ref{misner_sharp_m'}):
\begin{equation}
\frac{\partial}{\alpha \partial t}\left[ \Gamma \left( 1+e\right)
+u\frac{q}{\rho}\right]  =
-\frac{1}{\alpha }\frac{\partial }{\partial a}\left[ 4\pi 
r^{2}\alpha \left( up+\Gamma q\right) \right] .\label{eq_allenergy}
\end{equation}
 This immediately leads to the conservation equation 
(\ref{eq_total_energy}) for the total energy.
A tedious but straightforward calculation based on Eqs. 
(\ref{eq_allenergy}),
(\ref{eq_radius_space_deriv})-(\ref{eq_velocity_time_derivative}),
and (\ref{eq_phi_space_deriv}) finally leads to the momentum equation 
(\ref{eq_momentum}),
which completes the set of conservation equations

\begin{eqnarray}
\frac{\partial}{\partial t}\left[ \frac{1}{D}\right]  & = & \frac{\partial 
}{\partial a}\left[ 4\pi r^{2}\alpha u\right] \label{eq_continuity} \\
\frac{\partial \tau }{\partial t} & = & -\frac{\partial }{\partial a}\left[ 4\pi 
r^{2}\alpha \left( up+\Gamma q\right) \right] \label{eq_total_energy} 
\\
\frac{\partial S}{\partial t} & = & -\frac{\partial }{\partial a}\left[ 4\pi 
r^{2}\alpha \left( \Gamma p+uq\right) \right] \nonumber \\
 & - & \frac{\alpha }{r}\left[ \left( 1+e+\frac{3p}{\rho }\right) 
\frac{m}{r}+\frac{8\pi r^{2}}{\rho }\left( \rho \left( 1+e\right) 
p-q^{2}\right) -\frac{2p}{\rho }\right] \label{eq_momentum} \\
\frac{\partial V}{\partial a} & = & 
\frac{1}{D}\label{eq_volume_gradient} \\
\frac{\partial m}{\partial a} & = & 1+\tau \label{eq_mass_gradient} \\
\frac{\partial}{\partial t}\left[ \frac{1}{4\pi r^{2}\rho }\frac{q}{\rho }\right]  & 
= & -\left( 1+e\right) \frac{\partial \alpha }{\partial 
a}-\frac{1}{\rho }\frac{\partial }{\partial a}\left[ \alpha p\right] 
.\label{eq_lapse_gradient} 
\end{eqnarray}
The time derivative in the last equation usually is small,
so that this equation rather acts as a constraint on the lapse
function \( \alpha \). This equation is derived in the Appendix
from the space-space component of the four-divergence of the
stress-energy tensor. The constraints (\ref{eq_volume_gradient})
and (\ref{eq_mass_gradient}) explain themselves in analogy
to the Newtonian limit, where the first becomes the definition
of the rest mass density and the second the Poisson equation for
the gravitational potential. This set of conservative equations
is fundamental for the following discussion.
We will first point out their relation to the general relativistic 
jump conditions at a shock front and then derive a consistent
incorporation of artificial viscosity in Eqs.
(\ref{eq_continuity})-(\ref{eq_lapse_gradient}). In a second and third
part, we investigate the general relativistic Boltzmann equation with
two different natural descriptions of the momentum phase space for the
radiation particles. Both formulations will show interesting
relations to the above-mentioned conservation laws.


\section{Shock Waves and Artificial 
Viscosity}\label{section_artificial_viscosity}

The jump conditions across a shock front reflect the governing 
conservation laws and can directly be determined from the latter:
In an isolated shock wave, we assume a stable physical state at
the left-hand side (subscript \( L \)) and at the right-hand side (subscript 
\( R\)). The direction of shock propagation will not play a role
in this derivation. A conservation law can then be integrated over an 
infinitesimal range around the shock front,
containing rest mass \( \Delta a_{L} \) on the left-hand side
and \( \Delta a_{R} \) on the right-hand side of the shock at position
\( a_{s} \). For example, the 
continuity equation (\ref{eq_continuity}) leads to
\[
\frac{d}{dt} \left( \frac{\Delta a_{L}}{D_{L}} + \frac{\Delta 
a_{R}}{D_{R}} \right) =
4\pi r_{R}^{2}\alpha_{R}u_{R} - 4\pi r_{L}^{2}\alpha_{L}u_{L}.
\]
As the shock moves with \( da_{s}/dt = da_{L}/dt = -da_{R}/dt \) 
through rest mass the first jump condition reads
\begin{equation}
\left[ \frac{da_{s}}{dt}\frac{1}{D }+4\pi r^{2}u\alpha 
\right]  = 0.\label{eq_jump_continuity}
\end{equation}
The brackets denote the difference of the enclosed expression 
evaluated on both sides of the shock. With an analogous argument
applied to the energy conservation equation, the general
relativistic jump condition in spherical symmetry with energy
transport included reads
\begin{equation}
\left[ \frac{da_{s}}{dt}\tau -4\pi r^{2}\alpha \left( up+\Gamma q\right)
\right]  = 0. \label{eq_jump_energy}
\end{equation}
At this point one might guess the correct jump conditions for
the momentum conservation from Eq. (\ref{eq_momentum}) - in spite
of the complications introduced by the nonvanishing source term.
A rigorous derivation along the lines given by May and White
\cite{May_White_67} indeed leads to
\begin{equation}
\left[ \frac{da_{s}}{dt}S -4\pi r^{2}\alpha \left( \Gamma p + uq\right)
\right]  = 0.\label{eq_jump_momentum}
\end{equation}
Of course, with \( q=0 \), one immediately recovers the jump
conditions for pure hydrodynamics found in Ref. \cite{May_White_67}.

Although ideal hydrodynamics allows discontinuities in the physical state
across a shock front, there are many numerical schemes that cannot handle 
infinite gradients
in the velocity, density, or temperature profiles. A well-known 
solution to this problem is
the introduction of an artificial viscosity. The artificial viscosity 
limits the
gradient of the profiles and broadens the shock, which then becomes
spread over several
grid zones. With a conservative formulation, the dynamics of these 
asymptotically
incompressible zones becomes determined by conservation of volume, energy,
and momentum flux
across them. This guarantees the conservation of 
these quantities
over the whole shock as required by the jump conditions.
Older schemes use artificial viscosity in a form proposed by Von 
Neumann and Richtmyer \cite{VonNeumann_Richtmyer_50}.

In spherical symmetry, the formulation of artificial viscosity has to 
be chosen
with care in order to avoid systematic artificial heating during
homologous compression.
An excellent solution has been found by Tscharnuter and Winkler 
\cite{Tscharnuter_Winkler_79}
for nonrelativistic hydrodynamics. We extend this approach to 
general relativistic hydrodynamics and define the viscous tensor
\begin{eqnarray*}
  Q_{\alpha \beta } & = &
    \Delta l^{2}\rho u_{\; ;\mu }^{\mu }
      \left[ \varepsilon _{\alpha \beta }-\frac{1}{3}u_{\; ;\mu 
      }^{\mu }P_{\alpha \beta }\right]
    \qquad \mbox{if}\quad u^{\mu }_{\; ;\mu }<0, \\
    & = & 0 \qquad \mbox{otherwise}; \\
  \varepsilon _{\alpha \beta } & = & \frac{1}{2}
    \left( u_{\alpha ;\mu }P_{\; \beta }^{\mu }+u_{\beta ;\mu 
    }P_{\; \alpha }^{\mu }\right) ,
\end{eqnarray*}
where \( P_{\alpha \beta }=u_{\alpha }u_{\beta }+g_{\alpha \beta } \) 
is the
projection operator onto the three-space orthogonal to the fluid 
four-velocity \( u^{\alpha } \).
The artificial viscosity is based on physical viscosity and is chosen 
according
to the standard shear viscosity (Misner et al. 
\cite{Misner_Thorne_Wheeler_73},
exercise (22.6)). It is weighted with a variable coefficient 
that reflects the
local density and compression. The length scale \( \Delta l \) 
sets the order of magnitude of the desired shock width. 

First we note that the artificial viscosity tensor is traceless
and that
in our comoving coordinates its nonvanishing components are:
\begin{eqnarray}
Q^{a}_{\; a} & = & -2Q^{\vartheta }_{\; \vartheta }=-2Q^{\varphi 
}_{\; \varphi }=Q,\nonumber \\
Q & = &
\Delta l^{2}\rho \mbox {div}(u)\left[ \frac{\partial u}{\partial 
r}-\frac{1}{3}\mbox {div}(u)\right]  \qquad \mbox {if\quad div}(u)<0,
\nonumber\\
& = & 0 \qquad \mbox {otherwise};
\label{eq_definition_viscous_pressure} \\
\mbox {div}(u) & = & \frac{\partial }{\partial V}\left( 4\pi 
r^{2}u\right) .\nonumber
\end{eqnarray}
This viscous tensor is included in the derivation of the Einstein 
equations in spherical symmetry in the Appendix.
We can check the behavior of viscous heating in the energy equation 
(\ref{eq_internal_energy_time_deriv}):
\begin{eqnarray*}
\frac{\partial e}{\alpha \partial t} & = & \left( \frac{u}{r}-\frac{\partial 
u}{\partial r}\right) \frac{Q}{\rho }\\
 & = & -\frac{3}{2}\left[ \frac{\partial u}{\partial 
r}-\frac{1}{3}\mbox {div}(u)\right] \frac{Q}{r}.
\end{eqnarray*}
The first expression shows that the viscous heating vanishes in the 
case of
homologous compression \( (u/r=\partial u/\partial r) \). This is due 
to the
nonisotropy of the viscous pressure: With homologous compression, the 
work done
on a fluid element by radial compression against \( Q_{\: a}^{a} \) 
would heat
the fluid element. But the homologous compression comes together with 
a simultaneous
compression in \( \vartheta  \) and \( \varphi  \) direction. By 
choosing
the viscous pressure components to have negative sign in these 
directions, the
net heat production vanishes. The second expression together with Eq. 
(\ref{eq_definition_viscous_pressure})
shows that viscous heating always has positive sign. In a
conservative formulation,
the viscosity affects the equation for the total energy evolution 
(\ref{eq_total_energy}),
the momentum evolution (\ref{eq_momentum}), and the constraint for 
the lapse
function (\ref{eq_lapse_gradient}). These equations then become
\begin{eqnarray*}
\frac{\partial\tau }{\partial t} & = & -\frac{\partial }{\partial a}\left[ 4\pi 
r^{2}\alpha \left( u\left( p+Q\right) +\Gamma q\right) \right] \\
\frac{\partial S}{\partial t} & = & -\frac{\partial }{\partial a}\left[ 4\pi 
r^{2}\alpha \left( \Gamma \left( p+Q\right) +uq\right) \right] \\
 & - & \frac{\alpha }{r}\left[ \left( 1+e+\frac{3\left( p-Q\right) 
}{\rho }\right) \frac{m}{r}+\frac{8\pi r^{2}}{\rho }\left( \rho 
\left( 1+e\right) \left( p+Q\right) -q^{2}\right) -\frac{2p}{\rho 
}+\frac{Q}{\rho }\right] \\
\frac{\partial}{\partial t}\left[ \frac{1}{4\pi r^{2}\rho }\frac{q}{\rho }\right]  & 
= & -\left( 1+e\right) \frac{\partial \alpha }{\partial 
a}-\frac{1}{\rho }\frac{\partial }{\partial a}\left[ \alpha p\right] 
-\frac{1}{V\rho }\frac{\partial }{\partial a}\left[ V\alpha Q\right] .
\end{eqnarray*}


\section{Boltzmann Radiation 
Transport\label{chapter_radiation_transport}}

\subsection{General relativistic view}

The general relativistic Boltzmann equation in comoving coordinates 
was derived
by Lindquist \cite{Lindquist_66}. Radiation that is not necessarily in 
equilibrium with
the matter is described by a distribution function \( f \):
\[
dN=f(x,p)\left( -p_{\mu }u^{\mu }\right) d\tau dP.\]
The volume element \( d\tau \) is crossed at four-location \( x \) by \( 
dN \)
radiation particle world lines with four-momenta \( p \) in the range 
\( dP \)
while the observer moves with four-velocity \( u^{\mu } \). Lindquist 
\cite{Lindquist_66}
shows that this definition makes the distribution function Lorentz 
invariant. We measure the particle four-momentum
in a comoving orthonormal frame (\ref{eq_orthonormal_basis}),
with the components
\begin{equation}
\label{eq_neutrino_coordinates}
p^{a}=p\cos \theta ,\; p^{\vartheta}=p\sin \theta \cos \phi ,\;
p^{\varphi}=p\sin \theta \sin \phi .
\end{equation}
The absolute value, \( p \), of the three-momentum can be
determined from the scalar particle energy
\( E=-p_{\mu}u^{\mu}=p^{t}=\sqrt{p^{2}+m^{2}} \)
and the rest mass
(the mass of radiation particles is assumed to be zero in this 
paper).
The direction of the particle three momentum is specified by
the angle cosine \( \mu =\cos \theta \)
to the radial direction.
In spherical symmetry, the distribution function does not depend on 
the azimuth angle \( \phi  \).
Thus, the particle distribution function
depends on four arguments
\[
dN=f(t,a,\mu ,E)E^{2}dEd\mu \frac{dV}{\Gamma}.\]
The Boltzmann equation for metric (\ref{eq_comoving_metric}) then
reads (Lindquist \cite{Lindquist_66}, Eq. 
(3.7))
\begin{eqnarray}
\frac{1}{\alpha }\frac{\partial f}{\partial t} & + & \mu \frac{\Gamma}
{r'}\frac{\partial f}{\partial a}+\frac{\Gamma }{r}\left( 1-\mu 
^{2}\right) \frac{\partial f}{\partial \mu }\nonumber \\
 & - & \left[ \Gamma \frac{\partial \Phi }{\partial r}+\mu 
\frac{1}{\alpha }\frac{\partial \Lambda }{\partial t}-\mu \frac{u}{r}\right] \left( 
1-\mu ^{2}\right) \frac{\partial f}{\partial \mu }\nonumber \\
 & - & \left[ \mu \Gamma \frac{\partial \Phi }{\partial r}+\mu 
^{2}\frac{1}{\alpha }\frac{\partial \Lambda }{\partial t}+\left( 1-\mu ^{2}\right) 
\frac{u}{r}\right] E\frac{\partial f}{\partial E}=j+\chi 
f,\label{eq_boltzmann_lindquist} 
\end{eqnarray}
with \( \Lambda =\log (r'/\Gamma ) \). The right-hand side of the 
equation
is the collision term that describes changes in the particle
distribution function
due to interactions with matter. It is represented here by an 
emissivity
\( j \) and an absorptivity \( \chi  \).
Since the independent space coordinate is
enclosed rest mass, it is convenient to introduce the specific 
distribution
function \( F(t,a,\mu ,E)=f(t,a,\mu ,E)/\rho  \). With the help of 
the relations
in the Appendix, the Boltzmann equation 
(\ref{eq_boltzmann_lindquist}) can
be rewritten in the conservative form (Mezzacappa and Matzner 
\cite{Mezzacappa_Matzner_89},
Yamada et al. \cite{Yamada_Janka_Suzuki_99})
\begin{eqnarray}
\frac{\partial F}{\alpha \partial t} & + & \frac{\mu }{\alpha }\frac{\partial 
}{\partial a}\left[ 4\pi r^{2}\alpha \rho F\right] +\Gamma \left( 
\frac{1}{r}-\frac{1}{\alpha }\frac{\partial \alpha }{\partial 
r}\right) \frac{\partial }{\partial \mu }\left[ \left( 1-\mu 
^{2}\right) F\right] \nonumber \\
 & + & \left( \frac{\partial \ln \rho }{\alpha \partial t}+\frac{3u}{r}\right) 
\frac{\partial }{\partial \mu }\left[ \mu \left( 1-\mu ^{2}\right) 
F\right] \nonumber \\
 & + & \left( \mu ^{2}\left( \frac{\partial\ln \rho }{\alpha 
\partial t}+\frac{3u}{r}\right) -\frac{u}{r}-\mu \Gamma \frac{1}{\alpha 
}\frac{\partial \alpha }{\partial r}\right) 
\frac{1}{E^{2}}\frac{\partial }{\partial E}\left[ E^{3}F\right] 
=\frac{j}{\rho }+\chi F.\label{eq_relativistic_boltzmann} 
\end{eqnarray}
The first term is the temporal change of the particle distribution 
function.
The second term counts the particles that are propagating into 
or out of an infinitesimal mass shell. The third term
corrects for the change in the local propagation angle \( \mu  \) when 
the particle
moves to another radius. The curved particle trajectory in general 
relativity is accounted for by the term 
proportional to the gradient of the gravitational potential 
\[
\frac{1}{\alpha }\frac{\partial \alpha }{\partial 
r}=\frac{\partial \Phi }{\partial r}.
\]
The term on the second line accounts for angular aberration: When the 
comoving observers change radius, the corresponding relabeling
of particle propagation angles also adds a correction to the
in-out flow balance for the particle distribution function taken
at constant \( \mu \). The same is true for the frequency shift
described in the third line: The first two terms in the parentheses
account for the Doppler shift caused by comoving observers. The
relativistic term \( \mu \Gamma \partial \Phi /\partial r \)
describes the redshift or blueshift in the particle energy that
applies when the particles have a velocity component in the
radial direction (\( \mu \neq 0 \))
and therefore change their position in the gravitational potential.

The integration of the Boltzmann equation over momentum
space, spanned by the propagation angle and particle energy, is 
supposed to
reproduce the conservation laws for particle number and energy as 
stated
in Eqs. (\ref{eq_continuity}) and (\ref{eq_total_energy}).
We define \( H^{N} \) and \( G^{N} \) to represent the 
zeroth and first \( \mu \) moments of the distribution function:
\begin{eqnarray*}
H^{N} & = & \int ^{1}_{-1}\int ^{\infty }_{0}FE^{2}dEd\mu ,\\
G^{N} & = & \int ^{1}_{-1}\int ^{\infty }_{0}FE^{2}dE\mu d\mu .
\end{eqnarray*}
Integration of Eq. (\ref{eq_relativistic_boltzmann}) over momentum
space gives
the following evolution equation in terms of these moments:
\begin{equation}
\label{eq_neutrino_number_conservation}
\frac{\partial H^{N}}{\partial t}+\frac{\partial }{\partial a}\left[ 4\pi r^{2}\alpha 
\rho G^{N}\right] -\alpha \int \frac{j}{\rho }E^{2}dEd\mu +\alpha 
\int \chi FE^{2}dEd\mu =0.
\end{equation}
The aberration and frequency shift terms do not contribute because \( 
\left( 1-\mu ^{2}\right)  \)
vanishes at \( \mu =\pm 1 \) and \( E^{3}F \) is zero for \( E=0 \) 
as well
as for \( E=\infty  \). The nature of Eq. 
(\ref{eq_neutrino_number_conservation})
is a continuity equation analogous to Eq. (\ref{eq_continuity}), 
extended by
source and absorption terms for the radiation particles.
One more integration over the rest mass \( a \)
from the center of the star to its surface finally gives an equation 
for the
total radiation particle number variation as a function of integrated
emission and absorption.

Slightly less straightforward is the reproduction of the energy 
conservation.
Defining the energy moments as
\begin{eqnarray*}
H^{E} & = & \int FE^{3}dEd\mu \\
G^{E} & = & \int FE^{3}dE\mu d\mu = q/\rho\\
P^{E} & = & \int FE^{3}dE\mu ^{2}d\mu ,
\end{eqnarray*}
the corresponding integration of the Boltzmann equation 
(\ref{eq_relativistic_boltzmann})
first leads to the radiation energy equation and the radiation 
momentum equation
\begin{eqnarray*}
\frac{\partial H^{E}}{\partial t} & + & \frac{\partial }{\partial a}\left[ 4\pi 
r^{2}\alpha \rho G^{E}\right] +\frac{\alpha u}{r}\left( 
H^{E}-P^{E}\right) -\left( \frac{\partial \ln \rho }{\partial t}+\frac{2\alpha 
u}{r}\right) P^{E}\\
 & + & \Gamma \frac{\partial \alpha }{\partial r}G^{E}-\alpha \int 
\frac{j}{\rho }E^{3}dEd\mu +\alpha \int \chi FE^{3}dEd\mu =0,\\
\frac{\partial G^{E}}{\partial t} & + & \frac{\partial }{\partial a}\left[ 4\pi 
r^{2}\alpha \rho P^{E}\right] -\left( \frac{\alpha \Gamma }{r}-\Gamma 
\frac{\partial \alpha }{\partial r}\right) \left( H^{E}-P^{E}\right) 
-\left( \frac{\partial \ln \rho }{\partial t}+\frac{2\alpha u}{r}\right) G^{E}\\
 & + & \Gamma \frac{\partial \alpha }{\partial r}P^{E}+\alpha \int 
\chi FE^{3}dE\mu d\mu =0.
\end{eqnarray*}
Having Eq. (\ref{eq_allenergy}) in mind, we construct, in analogy
to \( \Gamma (1+e) + uq/\rho \), the specific radiation
energy \( \Gamma H^{E}+uG^{E} \) and investigate its time derivative. 
After a fair amount of algebra and the use of relations from the Appendix,
most of the contributions cancel and one is left with the expected
result
\begin{eqnarray}
0 & = & \frac{\partial}{\partial t}\left( \Gamma H^{E}+uG^{E}\right) +\frac{\partial 
}{\partial a}\left[ 4\pi r^{2}\alpha \rho \left( uP^{E}+\Gamma 
G^{E}\right) \right] \nonumber \\
 & - & \alpha \Gamma \int \frac{j}{\rho }E^{3}dEd\mu +\alpha \Gamma 
\int \chi FE^{3}dEd\mu +\alpha u\int \chi FE^{3}dE\mu d\mu 
.\label{eq_radiation_energy_conservation} 
\end{eqnarray}
The time derivative contains the conserved energy; the second term 
describes
the change of energy in the fluid element due to surface work by the 
radiation pressure,
\( P^{E} \), and in-out flow of radiation at luminosity
\( L=4\pi r^{2}\rho G^{E} \). Note that
\( (uP^E + \Gamma G^E )\), being the first \(\mu\) moment of
\( (\Gamma H^E + uG^E )\), indeed is
the flux of the conserved quantity at the zone boundary.
The same feature is realized in the radiation momentum equation
(\ref{eq_momentum}): \( \partial /\partial t (uH^E + \Gamma G^E ) =
- \partial /\partial a [4\pi r^2\alpha\rho (\Gamma P^E + u G^E)]
+ \ldots \). It underlines the natural appearance of this formulation.
The source terms in Eq. (\ref{eq_radiation_energy_conservation})
describe the energy exchange
with matter by emission, absorption, and work due to particle stress. 
Thus, by splitting the internal energy \( e=\widetilde{e}+H^{E} \) 
and pressure
\( p=\widetilde{p}+P^{E} \) into contributions from the matter
stress-energy
tensor and the radiation stress-energy tensor, the exact energy 
conservation
equation can finally be written concisely as
\begin{eqnarray}
\frac{\partial}{\partial t}\left[ \Gamma \widetilde{e}+\frac{2}{\Gamma +1}\left( 
\frac{1}{2}u^{2}-\frac{m}{r}\right) +\Gamma H^{E}+uG^{E}\right]  &  & 
\nonumber \\
+\frac{\partial }{\partial a}\left[ 4\pi r^{2}\alpha 
u\widetilde{p}+4\pi r^{2}\alpha \rho \left( uP^{E}+\Gamma 
G^{E}\right) \right]  & = & 
0.\label{eq_hydro+radiation_energy_conservation} 
\end{eqnarray}


\subsection{Order \( v/c \) limit}

The \( O(v/c) \) Boltzmann equation was derived by Castor 
\cite{Castor_72}.
One can also obtain it by directly dropping the higher order terms 
from Eq.
(\ref{eq_relativistic_boltzmann}), i.e., by setting \( \alpha = \Gamma
= 1 = const. \) and replacing \( u \) by the nonrelativistic
velocity \( v \):
\begin{eqnarray}
\frac{\partial F}{\partial t} & + & \mu \frac{\partial }{\partial a}\left[ 4\pi 
r^{2}\rho F\right] +\frac{1}{r}\frac{\partial }{\partial \mu }\left[ 
\left( 1-\mu ^{2}\right) F\right] \nonumber \\
 & + & \left( \frac{\partial \ln \rho }{\partial t}+\frac{3v}{r}\right) 
\frac{\partial }{\partial \mu }\left[ \mu \left( 1-\mu ^{2}\right) 
F\right] \nonumber \\
 & + & \left( \mu ^{2}\left( \frac{\partial \ln \rho 
}{\partial t}+\frac{3v}{r}\right) -\frac{v}{r}\right) 
\frac{1}{E^{2}}\frac{\partial }{\partial E}\left[ E^{3}F\right] 
=\frac{j}{\rho }+\chi F.\label{eq_ovc_boltzmann} 
\end{eqnarray}
The conservation properties of the \( O(v/c) \) Boltzmann equation
become apparent in terms of the moments that are directly derived from
Eq. (\ref{eq_ovc_boltzmann}):
\begin{eqnarray*}
\frac{\partial H^{E}}{\partial t} & + & \frac{\partial }{\partial a}\left[ 4\pi 
r^{2}\rho G^{E}\right] +\frac{v}{r}\left( H^{E}-P^{E}\right) -\left( 
\frac{\partial \ln \rho }{\partial t}+\frac{2v}{r}\right) P^{E}\\
 & - & \int \frac{j}{\rho }E^{3}dEd\mu +\int \chi FE^{3}dEd\mu =0,\\
\frac{\partial G^{E}}{\partial t} & + & \frac{\partial }{\partial a}\left[ 4\pi 
r^{2}\rho P^{E}\right] -\frac{1}{r}\left( H^{E}-P^{E}\right) -\left( 
\frac{\partial \ln \rho }{\partial t}+\frac{2v}{r}\right) G^{E}\\
 & + & \int \chi FE^{3}dE\mu d\mu =0.
\end{eqnarray*}
When we construct the conserved specific radiation energy
\( \left( \Gamma H^{E}+uG^{E}\right)  \) in the \( O(v/c) \) limit,
we first keep all
terms that arise from the \( O(v/c) \) Boltzmann equation and 
obtain
\begin{eqnarray}
0 & = & \frac{\partial}{\partial t}\left( H^{E}+vG^{E}\right) +\frac{\partial 
}{\partial a}\left[ 4\pi r^{2}\rho \left( vP^{E}+G^{E}\right) \right] 
\nonumber \\
 & - & \int \frac{j}{\rho }E^{3}dEd\mu +\int \chi FE^{3}dEd\mu +v\int 
\chi FE^{3}dE\mu d\mu \nonumber \\
 & - & \frac{1}{4\pi r^{2}\rho }\frac{\partial}{\partial t}\left( 4\pi r^{2}\rho
v\right) G^{E}.\label{eq_ovc_radiation_energy_conservation} 
\end{eqnarray}
Note that the first order terms
\[
\frac{v}{r}\left( H^{E}-P^{E}\right) -\left(
\frac{\partial \ln \rho }{\partial t}+\frac{2v}{r}\right) P^{E}
\]
arising from frequency shift and angular aberration
in the radiation energy equation cancel with the zeroth order 
terms in the
radiation momentum equation after the latter are multiplied by
the fluid velocity. In a numerical implementation, the finite
differencing of the Boltzmann equation has to guarantee the
same cancellation in order to achieve \( O(v/c) \) energy
conservation. The nonconservative term
found in the third line in Eq.
(\ref{eq_ovc_radiation_energy_conservation})
is of second order in \( v/c \).


\section{Approximate Tangent Ray 
Description\label{section_boltzmann_phasespace} }

\subsection{General relativistic view}

The description of the particle four-momenta in the comoving
frame eases the numerical implementation of the collision
term. The angle dependent integrals over reaction cross sections of
particles with matter are readily evaluated if the
target is at rest in the comoving frame. On the other
hand, the left-hand side of the Boltzmann equation
(\ref{eq_relativistic_boltzmann}) has
to take into account the Lorentz transformation between 
adjacent comoving observers in the description of the
changes in the particle distribution function due to
propagation. An adequat discretization of the
numerous correction terms with partial derivatives
is, in spherical symmetry, a resolved challenge (Mezzacappa
and Bruenn \cite{Mezzacappa_Bruenn_93a},
Liebend\"orfer \cite{Liebendoerfer_00}). However, this
does not prevent consideration of other interesting options.
Here we would like to look into a description of the
momentum phase space in variables that remain constant
along a propagation path in the absence of interactions.
Particles change their coordinates in the momentum phase
space only due to collisions, evaluated on the right-hand
side of the Boltzmann equation. The left-hand side
does not require corrections involving partial
derivatives with respect to the coordinates of the
momentum phase space. The drawback is given by the unavoidable
need of transformations to relate the cross sections
in the fluid rest frame to the distribution function in
the chosen momentum phase space description.
This idea was investigated in polar slicing and the radial 
gauge by Schinder and Bludman \cite{Schinder_Bludman_89}.
We explore a similar ansatz in orthogonal comoving
coordinates in this section.

A geodesic in a static spherically symmetric space-time can uniquely be 
described
by an impact parameter \( b \) and a particle energy \( \varepsilon  
\) at
infinity. Let us start with a Schwarzschild metric

\begin{equation}
\label{eq_schwarzschild_metric}
ds^{2}=-\Gamma ^{2}_{S}dt^{2}_{S}+\Gamma ^{-2}_{S}dr^{2}+r^{2}\left( 
d\vartheta +\sin ^{2}\vartheta d\varphi ^{2}\right) ,
\end{equation}
where we add a subscript \( S \) to quantities that could be 
confused
with corresponding quantities in the comoving frame (e.g.,
\( \Gamma _{S}=\sqrt{1-2m/r} \)).
The trajectory of free propagation in the plane of 
constant \( \varphi  \)
with energy \( E_{S} \) is related to the impact parameter \( b \) 
and the
energy at infinity \( \varepsilon  \) as (Misner et al. 
\cite{Misner_Thorne_Wheeler_73},
Eqs. (25.55) and (25.18/19))
\begin{eqnarray}
\left( \frac{1}{r}\frac{dr}{d\vartheta }\right) ^{2}+1-\frac{2m}{r} & 
= & \frac{r^{2}}{b^{2}} \label{eq_static_trajectory} \\
\Gamma _{S}E_{S} & = & \varepsilon .\label{eq_static_energy}
\end{eqnarray}
The relation between the particle propagation in the \( r \)- and \( 
\vartheta \) directions
is given by the angle \( \theta  \) the particle trajectory makes
with the outward radial direction:
\begin{equation}
\tan \theta =\frac{\sqrt{1-\mu ^{2}_{S}}}{\mu _{S}}=\frac{rd\vartheta 
}{\Gamma ^{-1}_{S}dr}.\label{eq_static_angle}
\end{equation}
Equations (\ref{eq_static_trajectory}) and (\ref{eq_static_angle})
are solved for the relation between the particle
angle, \( \mu_S \) and the impact parameter:
\begin{equation}
b=\frac{r}{\Gamma _{S}}\sqrt{1-\mu ^{2}_{S}.}
\label{eq_static_impact}
\end{equation}

Our next step is to transform \( \mu _{S} \) and \( E_{S} \) 
into the particle angle, \( \mu  \), and energy, \( E \), measured by the 
comoving observers.
From the metrics (\ref{eq_comoving_metric}) and (\ref{eq_schwarzschild_metric})
one obtains the relations
\begin{eqnarray}
\frac{\partial t_{S}}{\partial t} & = &
\frac{\alpha \Gamma}{\Gamma ^{2}_{S}} \nonumber \\
\frac{\partial t_{S}}{\partial a} & = &
 \frac{r'}{\Gamma} \frac{u}{\Gamma ^{2}_{S}}
\label{eq_lorentz_transformation}
\end{eqnarray}
for the Lorentz transformation between the Schwarzschild- and
comoving-coordinate time. The Lorentz transformation also
links the particle four-momenta in these two coordinate bases:
\begin{eqnarray}
p^{\mu }_{S} & = & \left( \Gamma ^{-1}_{S},\Gamma _{S}\mu 
_{S},\frac{1}{r}\sqrt{1-\mu ^{2}_{S}},0\right) E_{S} \nonumber \\
p^{\mu } & = & \left( \alpha ^{-1},\frac{\Gamma }{r'}\mu 
,\frac{1}{r}\sqrt{1-\mu ^{2}},0\right) E.\label{eq_momenta_comparison}
\end{eqnarray}
From the application of transformation (\ref{eq_lorentz_transformation})
to the four-momenta in Eqs. (\ref{eq_momenta_comparison}), we extract
the transformation of the particle propagation angle and energy:
\begin{eqnarray*}
\frac{1-\mu ^{2}_{S}}{\Gamma _{S}^{2}} & = & \frac{1-\mu ^{2}}
{\left( \Gamma +u\mu \right) ^{2}} \\
\Gamma _{S}E_{S} & = & \left( \Gamma +u\mu \right) E. 
\end{eqnarray*}
Finally, Eqs. (\ref{eq_static_impact}) and (\ref{eq_static_energy})
become, in terms of comoving frame variables,
\begin{eqnarray}
b & = & r\frac{\sqrt{1-\mu ^{2}}}{\Gamma +u\mu }\label{eq_impact}\\
\varepsilon  & = & \left( \Gamma +u\mu \right) E.\label{eq_einfinity}
\end{eqnarray}
These useful relations link the local particle angles and energies
measured by comoving observers at different radii and with different
velocities to the particle impact parameter and energy at infinity.
Additionally, these relations provide insight into the nature
of the energy conservation
equation (\ref{eq_radiation_energy_conservation}). The conserved 
quantity under the time derivative
\[
\Gamma H^{E}+uG^{E}=\int \left( \Gamma +u\mu \right) FE^{3}dEd\mu 
=\int \varepsilon FE^{2}dEd\mu\]
is the total radiation energy \emph{at infinity}
expressed in terms of the comoving frame radiation energy
and momentum. This also makes clear the following important
point: The locally observed radiation quantities
must be transformed to a common observation point (e.g., infinity) 
before they can be integrated to define a conserved quantity. 

Let us recall that the particle trajectories along constant
impact parameter, \( b \), and constant energy at infinity,
\( \varepsilon \), were derived in static vacuum Schwarzschild
space-time surrounding a gravitational mass \( m \).
Nevertheless, we can span the momentum phase space in the
dynamical Boltzmann equation in comoving coordinates, Eq.
(\ref{eq_relativistic_boltzmann}), with the coordinates
\footnote{
Strictly speaking, the pair \( (b,\varepsilon) \) only
specifies a trajectory. The propagation direction on this
trajectory can, for example, be addressed with the convention
that a negative impact parameter denotes a propagation
with decreasing radius and a positive impact parameter 
a propagation with increasing radius on the same trajectory.
}
\( (b,\varepsilon) \)
instead of the comoving frame four-momenta \( (\mu,E) \).
If the static trajectories are a good approximation to the
trajectories on a dynamical background, we expect that the
momentum state of free particle propagation,
measured in \( (b,\varepsilon) \),
barely changes between adjacent comoving observers. This
would imply
that the terms involving partial derivatives with respect
to \( b \) and \( \varepsilon \) become small compared to the
time derivative of the distribution function and the
in-out flow term on the left-hand side of the exact
Boltzmann equation.

We replace the partial derivatives in the Boltzmann
equation by directional derivatives along the static particle
trajectory: We add an overbar to the distribution
function \( \overline{f}(t,a,b,\varepsilon )=f(t,a,\mu ,E) \) in 
order to indicate
that the partial derivatives are taken at constant impact parameter
\( b \) and energy at infinity \( \varepsilon  \) and make the
following ansatz:
\begin{equation}
C_{1}\frac{\partial\overline{f}}{\partial t}+C_{2}\frac{\partial 
\overline{f}}{\partial a}+C_{3}\frac{\partial \overline{f}}{\partial 
b}+C_{4}\frac{\partial \overline{f}}{\partial \varepsilon }=j+\chi 
\overline{f}.\label{eq_boltzmann_ansatz}
\end{equation}
The coefficients \( C_{i} \) are calculated by comparing the 
Boltzmann equation (\ref{eq_boltzmann_lindquist}) to
Eq. (\ref{eq_boltzmann_ansatz}). The relations (\ref{eq_impact})
and (\ref{eq_einfinity}) are used in a straightforward but
lengthy replacement of the partial derivatives
in Eq. (\ref{eq_boltzmann_lindquist}) by derivatives
with respect to impact parameter and energy at infinity.
The Boltzmann equation written in terms of these directional
derivatives reads
\begin{eqnarray}
\frac{1}{\alpha }\frac{\partial\overline{f}}{\partial t} & + & \mu \frac{\Gamma 
}{r'}\frac{\partial \overline{f}}{\partial a}
+ \frac{4\pi r}{\Gamma +u\mu }\left( \mu\rho \left( 1+e+\frac{p}{\rho }\right) 
- \left( 1+\mu^2 \right) q \right)
\nonumber \\
 & \times  & \left( -\varepsilon \frac{\partial 
\overline{f}}{\partial \varepsilon }+b\frac{\partial 
\overline{f}}{\partial b}\right) =j+\chi 
\overline{f}.\label{eq_boltzmann_phasespace} 
\end{eqnarray}
Its left-hand side (LHS) involves the first two ususal propagation terms
that relate the change in the distribution function to particle
in-out flow at the boundaries of
a mass element. The third term arises owing to the drift
of the particle location in the phase space from constant
\( (b,\varepsilon) \) in front of a non-vacuum background.
This general relativistic term is of order \( (G\rho r^2/c^2) \)
and vanishes in the vacuum. It is interesting to see
in the static limit that this term mainly arises from
the background presence of matter and not from its
dynamical motion. We might then neglect this \( O(G\rho r^2/c^2) \)
term in the exact equation (\ref{eq_boltzmann_phasespace})
and find to the very simple approximative general relativistic
Boltzmann equation in spherical symmetry:
\begin{equation}
\frac{1}{\alpha }\frac{\partial\overline{f}}{\partial t}+\mu \frac{\Gamma 
}{r'}\frac{\partial \overline{f}}{\partial a}=j+\chi \overline{f}.
\label{eq_boltzmann_approximate}
\end{equation}
This approximation is excellent in the proximity of compact
objects; it includes full gravitational redshift. It
might however fail within regions of extremely high energy
density. Moreover, we show in the next section
that its \( O(v/c) \) expansion is identical with the
full \( O(v/c) \) Boltzmann equation (\ref{eq_ovc_boltzmann}).
Equation (\ref{eq_boltzmann_approximate}) therefore also incorporates
an angular aberration and Doppler shift between the comoving
coordinate frames.
Its intuitive form simply reflects Lindquist's general starting point 
\cite{Lindquist_66}
(2.12/22), where the Boltzmann equation is written in terms of 
directional derivatives
of the distribution function along the phase flow.

\subsection{Order \( v/c \) limit\label{section_tangent_ovc}}
First, we multiply the left hand side of the \( O(v/c) \) Boltzmann
equation (\ref{eq_ovc_boltzmann}) by the density, \( \rho \),
express it in terms of the neutrino
distribution function \( f=\rho F\), and take the prefactors out
of the derivatives to obtain the formulation presented by
 Castor \cite{Castor_72}. With the continuity equation
\begin{equation}
-\frac{\partial \ln \rho }{\partial t}=\frac{v'}{r'}+\frac{2v}{r}
\end{equation}
we then find
\begin{equation}
LHS=\frac{\partial f}{\partial t}+\frac{\mu }{r'}\frac{\partial f}{\partial a}
+\left( 1-\mu ^{2}\right) \left[ \mu \left( \frac{v}{r}-\frac{v'}{r'}\right)\
+\frac{1}{r}\right]
 \frac{\partial f}{\partial \mu }-\left[ \mu ^{2}\frac{v'}{r'}
+\left( 1-\mu ^{2}\right) \frac{v}{r}\right] E\frac{\partial f}{\partial E}.
\label{eq_proof_ovc}
\end{equation}

On the other hand, we expand the approximate general relativistic
Boltzmann equation (\ref{eq_boltzmann_approximate}) to order \( v/c \)
and compare to Eq. (\ref{eq_proof_ovc}). Before the expansion, however,
we have to transform the directional derivatives back to partial
derivatives with respect to comoving frame energy, \( E \), and
angle, \( \mu \). In terms of an unspecified parameter \( \lambda \),
Eqs. (\ref{eq_impact}) and (\ref{eq_einfinity}) lead to the derivatives
of the comoving frame propagation angle and particle energy,
taken at constant impact parameter and energy at infinity:
\begin{eqnarray}
\left( \frac{\partial \mu }{\partial \lambda }\right) _{b,\varepsilon }
& = & \left( 1-\mu ^{2}\right) \frac{1+v\mu }{v+\mu }
\left( \frac{1}{r}\frac{\partial r}{\partial \lambda }
-\frac{\mu }{1+v\mu }\frac{\partial v}{\partial \lambda }\right)
\nonumber \\
\left( \frac{\partial E}{\partial \lambda }\right) _{b,\varepsilon }
& = & -\frac{E}{v+\mu }\left( \left( 1-\mu ^{2}\right) \frac{v}{r}
\frac{\partial r}{\partial \lambda }+\mu ^{2}
\frac{\partial v}{\partial \lambda }\right).
\label{eq_partial_relations}
\end{eqnarray}
The evaluation of the partial derivatives on the left hand
side of Eq. (\ref{eq_boltzmann_approximate}) is
straightforward with the relations (\ref{eq_partial_relations}).
We obtain
\begin{eqnarray}
LHS & = & \frac{\partial f}{\partial t}
+\frac{\partial f}{\partial \mu }\frac{\partial \mu }{\partial t}
+\frac{\partial f}{\partial E}\frac{\partial E}{\partial t}
+\frac{\mu }{r'}\frac{\partial f}{\partial a}
+\frac{\mu }{r'}\frac{\partial f}{\partial \mu }\frac{\partial \mu }{\partial a}
+\frac{\mu }{r'}\frac{\partial f}{\partial E}\frac{\partial E}{\partial a}
\nonumber \\
& = & \frac{\partial f}{\partial t}+\frac{\mu }{r'}
\frac{\partial f}{\partial a}+\left( 1-\mu ^{2}\right)
 \left[ \left( 1+v\mu \right) \frac{1}{r}-\frac{\mu }{1+v/\mu }
\frac{v'}{r'}\right] \frac{\partial f}{\partial \mu } \nonumber \\
& - & \left[ \left( 1-\mu ^{2}\right) \frac{v}{r}+\frac{\mu ^{2}}{1+v/\mu }
\frac{v'}{r'}\right] E\frac{\partial f}{\partial E}.
\label{eq_ovc_approximate}
\end{eqnarray}
Following Bruenn \cite{Bruenn_85}, we have already dropped the terms
involving the matter acceleration \( \partial v/\partial t \) in the
last step. If we further neglect the order \( v/c \) and higher order
terms, Eq. (\ref{eq_ovc_approximate}) reduces exactly to
Eq. (\ref{eq_proof_ovc}). We therefore conclude that the approximate
general relativistic Boltzmann equation (\ref{eq_boltzmann_approximate})
exactly includes and, moreover, extends the \( O(v/c) \) Boltzmann
equation in the presence of strong gravitational fields.


\section{Conclusion}

We provide the exact Einstein equations for radiation
hydrodynamics in spherical
symmetry, formulated in the comoving orthogonal coordinates 
introduced by Misner
and Sharp \cite{Misner_Sharp_64,Misner_Sharp_65}. 
We show that these equations
can be written in a concise and strictly conservative form. 
The general relativistic jump conditions at shock fronts
are derived under inclusion of the radiation energy flux and
radiation momentum flux.

The conservative formulation is ideally suited for
the application of numerous numerical schemes specifically designed
for the solution of partial differential equations in 
this form:
for example, adaptive grid techniques (Winkler et al. 
\cite{Winkler_Norman_Mihalas_84}),
Riemann solvers (Godunov \cite{Godunov_59}, Davis \cite{Davis_88}),
and high-resolution shock capturing schemes
(Romero et al. \cite{Romero_et_al_96},
Marti and M\"uller \cite{Marti_Mueller_99}).
Some of these schemes require artificial viscosity
for a numerical representation of shock waves. We
have generalized the tensor artificial viscosity ansatz of 
Tscharnuter and
Winkler \cite{Tscharnuter_Winkler_79} to general relativity. As in 
the nonrelativistic
case, viscous heating is shown to always have positive sign and to 
vanish during
homologous collapse.

In the second part of this paper, we focus on the general
relativistic Boltzmann equation in our chosen coordinates.
We confirm that, as
expected, the appropriate moments of the particle distribution
function, evolved according to the Boltzmann transport equation,
obey the conservation laws found in the first part of the paper.
In the \( O(v/c) \) limit, we identify leading terms that
require careful discretization in a finite difference
representation of the Boltzmann equation.
Angular aberration and frequency shift corrections arise
in the propagation part of the Boltzmann equation since
the time and space derivative of the particle distribution
function is taken at constant comoving frame four-momenta
in the particle momentum phase space.

In the last part, we try to recover the simple nature
of the left-hand side of the transport equation,
understanding it as a directional derivative along the
phase flow of the particles. The approximation of the
phase flow by geodesics in vacuum Schwarzschild space-time
leads first to the exact general relativistic Boltzmann
equation with the momentum phase space now described
in terms of particle impact parameter and energy at infinity.
The correction terms to the partial derivatives with
respect to time and space along vaccum geodesics are
of order \( (G\rho r^2/c^2) \) in this representation. They
are only due to the background matter distribution
neglected in the determination of the phase flow.
We show that neglecting these terms leads to a very
intuitive Boltzmann equation that is exact in the
\( O(v/c) \) limit and additionally includes
gravitational redshift. Although the numerical
implementation of this equation has to solve
other problems (for example, the radius dependent
range of valid impact parameters \( b = [r,\infty ] \)),
the exploration of this idea reveals the physical
interpretation of the conservation laws found in
the first and second parts: The conserved quantities
are those measured by an observer at infinity, but
expressed in terms of local quantities measured by
comoving observers.

After Wilson's \cite{Wilson_71} pioneering work, years of code
development by several groups (Mezzacappa and Bruenn
\cite{Mezzacappa_Bruenn_93a}, Yamada et al. \cite{Yamada_Janka_Suzuki_99},
Burrows et al. \cite{Burrows_et_al_00},
Liebend\"orfer \cite{Liebendoerfer_00},
Rampp \cite{Rampp_00}, Messer \cite{Messer_00}) have
led to Boltzmann solvers for neutrino transport in spherical
symmetry. Dynamical simulations of core collapse supernovae
have been presented in the Newtonian limit by
Mezzacappa et al. \cite{Mezzacappa_et_al_00} and
Rampp and Janka \cite{Rampp_Janka_00}. As general relativistic 
effects significantly
influence the postbounce evolution of a collapsed star (Bruenn et al. 
\cite{Bruenn_DeNisco_Mezzacappa_00}),
these codes have ultimately to be extended to include full general 
relativistic
radiation hydrodynamics. Such simulations were performed by 
Liebend\"orfer et al. \cite{Liebendoerfer_00,Liebendoerfer_et_al_00}
based on the equations discussed in the first and second parts,
under omission of neutrino back reaction.
Awareness of the conservation laws is essential in the simulation of 
supernovae because the observed explosion energy is two orders of
magnitude smaller than, for example, the core binding energy or
the radiation (neutrino) energy. It has to be 
excluded that an energy drift, owing to numerical inaccuracy in
energy conservation, seriously affects the equalized balance
and causes or prevents an explosion.


\section*{Acknowledgments}
We thank Jacob Fisker and
Jochen Peitz for helpful discussions, and Christian Cardall for
comments on the paper. In addition to the support of the
Swiss National Science Foundation under contract 20-53798.98,
M.L. is supported by the National Science
Foundation under contract AST-9877130,
A.M. is supported at the Oak Ridge National Laboratory, managed by
UT-Battelle, LLC, for the U.S. Department of Energy under contract
DE-AC05-00OR22725, and
F.-K.T. is supported in part by the Swiss National Science Foundation 
under contract 20-61822.00 and as a Visiting Distinguished Scientist
at the Oak Ridge National Laboratory. 


\appendix

\section{Derivation of the Einstein Equations in Spherical 
Symmetry}\label{appendix_einstein}

In this appendix, we derive the Einstein equations in comoving 
coordinates and
spherical symmetry. We follow closely the guidelines of exercise 
(14.16) in
Misner et al. \cite{Misner_Thorne_Wheeler_73}. In spherical symmetry,
we are
allowed to make the following ansatz for the metric:
\[
ds^{2}=-e^{2\Phi (a,t)}dt^{2}+e^{2\Lambda 
(a,t)}da^{2}+r^{2}(a,t)\left( d\vartheta ^{2}+\sin ^{2}(\vartheta 
)d\varphi ^{2}\right) .\]
The coordinates consist of an independent time coordinate \( t \) and 
a space
coordinate \( a. \) The areal radius \( r(a,t) \) is chosen such that
the area of the two-sphere,
\( d\vartheta^{2} + sin^{2}(\vartheta)d\varphi^{2} \), is \( 4\pi r^2 \).
We define the following orthonormal noncoordinate basis
\begin{eqnarray}
\omega ^{t} & = & e^{\Phi (a,t)}dt \nonumber \\
\omega ^{a} & = & e^{\Lambda (a,t)}da \nonumber \\
\omega ^{\vartheta } & = & r(a,t)d\vartheta \nonumber \\
\omega ^{\varphi } & = & r(a,t)\sin (\vartheta )d\varphi 
\label{eq_orthonormal_basis}
\end{eqnarray}
and calculate its exterior derivatives. From \( 0=d\omega ^{\mu 
}+\omega _{\: \nu }^{\mu }\wedge \omega ^{\nu } \)
(14.31a,b in Misner et al. \cite{Misner_Thorne_Wheeler_73}) one 
determines
the nonvanishing connections
\begin{eqnarray}
\omega _{ta} & = & -\Phi 'e^{-\Lambda }\omega ^{t}-\dot{\Lambda 
}e^{-\Phi }\omega ^{a}\nonumber \\
\omega _{t\vartheta } & = & -\frac{\dot{r}}{r}e^{-\Phi }\omega 
^{\vartheta }\nonumber \\
\omega _{t\varphi } & = & -\frac{\dot{r}}{r}e^{-\Phi }\omega 
^{\varphi }\nonumber \\
\omega _{a\vartheta } & = & -\frac{r'}{r}e^{-\Lambda }\omega 
^{\vartheta }\nonumber \\
\omega _{a\varphi } & = & -\frac{r'}{r}e^{-\Lambda }\omega ^{\varphi 
}\nonumber \\
\omega _{\vartheta \varphi } & = & -\frac{1}{r}\frac{\cos (\vartheta 
)}{\sin (\vartheta )}\omega ^{\varphi }.\label{eq_connection_forms} 
\end{eqnarray}
An overdot denotes the derivative with respect to coordinate time
\( t \) and a prime denotes the derivative with respect to the
spatial coordinate \( a \). The exterior derivatives
of these connections lead to the curvature tensor, and finally
the Einstein tensor:
\begin{eqnarray*}
G_{\: t}^{t} & = & -\left( F+2\overline{F}\right) \\
G_{\: a}^{t} & = & 2H\\
G_{\: a}^{a} & = & -\left( F+2\overline{E}\right) \\
G_{\: \vartheta }^{\vartheta } & = & -\left( 
\overline{E}+E+\overline{F}\right) \\
G_{\: \varphi }^{\varphi } & = & -\left( 
E+\overline{F}+\overline{E}\right) ,
\end{eqnarray*}
where
\begin{eqnarray*}
E & = & \left( -\Phi ''+\Phi '\Lambda '-\Phi ^{\prime 2}\right) 
e^{-2\Lambda }+\left( \ddot{\Lambda }-\dot{\Phi }\dot{\Lambda 
}+\dot{\Lambda }^{2}\right) e^{-2\Phi }\\
\overline{E} & = & \frac{1}{r}\left[ \left( \ddot{r}-\dot{r}\dot{\Phi 
}\right) e^{-2\Phi }-r'\Phi 'e^{-2\Lambda }\right] \\
H & = & \frac{1}{r}\left( \dot{r}'-\dot{r}\Phi '-r'\dot{\Lambda 
}\right) e^{-\left( \Phi +\Lambda \right) }\\
F & = & \frac{1}{r^{2}}\left( 1+\dot{r}^{2}e^{-2\Phi }-r^{\prime 
2}e^{-2\Lambda }\right) \\
\overline{F} & = & \frac{1}{r}\left[ \left( -r''+r'\Lambda '\right) 
e^{-2\Lambda }+\dot{r}\dot{\Lambda }e^{-2\Phi }\right] .
\end{eqnarray*}

In the following, we try to write the Einstein equations \( G=8\pi T 
\) as
concise as possible in terms of quantities that are defined in the 
restframe
of a fluid. The nonvanishing components of the stress-energy tensor 
of a fluid in its comoving frame are given in
Eq. (\ref{eq_stress_energy_tensor}). We include an addition,
\( Q \), to the diagonal component, representing artificial
viscosity as defined in Sec.
\ref{section_artificial_viscosity},
Eq. (\ref{eq_definition_viscous_pressure}):
\[
T^{aa} = p+Q, \qquad
T^{\vartheta \vartheta }=T^{\varphi \varphi } = p-\frac{1}{2}Q.
\]
First, we eliminate the exponentials in
the metric by substituting the lapse function \( \alpha  \)
for \( e^{\Phi} \) and the function \( r'/\Gamma  \) for
\( e^{\Lambda} \) in order
to retrieve notation (\ref{eq_comoving_metric}).
Then, we have to make sure that the yet unspecified space coordinate 
\( a \)
is attached to matter, i.e., that it is a Lagrangian (comoving)
coordinate. The rest mass between coordinate \( a_{0} \) 
and \( a_{1} \)
at a fixed coordinate time \( t \) is given from the metric by
\[
A(a_{0},a_{1})=\int ^{2\pi }_{0}\int ^{\pi }_{0}\int 
^{a_{1}}_{a_{0}}\rho \frac{r'}{\Gamma }da\: rd\vartheta \: r\sin 
(\vartheta )d\varphi =\int ^{a_{1}}_{a_{0}}\frac{4\pi r^{2}r'\rho 
}{\Gamma }da.\]
We tie the spatial coordinate to rest mass by requiring that \( 
A(a_{0},a_{1})=\int ^{a_{1}}_{a_{0}}da \)
for arbitrary boundaries \( a_{0} \), \( a_{1} \). This is equivalent 
to adopting the relation
\begin{equation}
\label{eq_radius_space_deriv}
r'=\frac{\Gamma }{4\pi r^{2}\rho }.
\end{equation}
Further, we define a ``velocity'' \( u=\dot{r}/\alpha  \) that 
describes
the change of areal radius with proper time of the comoving observer. 

From the nondiagonal component of the Einstein equations we derive
in three steps:
\begin{eqnarray}
\frac{1}{2}G^{ta}=H & = & \frac{\Gamma }{r}\left[ 
\frac{u'}{r'}-\frac{1}{\alpha }\left( 
\frac{\dot{r}'}{r'}-\frac{\dot{\Gamma }}{\Gamma }\right) \right] 
=4\pi q,\nonumber \\
\frac{1}{\alpha }\left( \frac{\dot{r}'}{r'}-\frac{\dot{\Gamma 
}}{\Gamma }\right)  & = & \frac{u'}{r'}-\frac{4\pi rq}{\Gamma 
},\label{eq_einstein_nodiagonal_component} \\
\frac{1}{\alpha }\frac{\dot{\Gamma }}{\Gamma } & = & u\frac{\Phi 
'}{r'}+\frac{4\pi rq}{\Gamma }.\label{eq_gamma_time_deriv} 
\end{eqnarray}
Equations (\ref{eq_radius_space_deriv}) and 
(\ref{eq_einstein_nodiagonal_component})
lead in three further steps to the evolution
equation for the rest mass density:
\begin{eqnarray}
\frac{\partial}{\alpha \partial t}\left( \frac{1}{\rho }\right)  & = & 
\frac{\partial}{\alpha \partial t}\left( \frac{4\pi r^{2}r'}{\Gamma }\right) 
\nonumber \\
 & = & \frac{4\pi r^{2}r'}{\Gamma }\left( 
\frac{2u}{r}+\frac{1}{\alpha }\left( 
\frac{\dot{r}'}{r'}-\frac{\dot{\Gamma }}{\Gamma }\right) \right) 
,\nonumber \\
& = & \frac{1}{\Gamma }\left[ \left( 4\pi r^{2}u\right) ^{\prime 
}-\frac{4\pi rq}{\rho }\right] .\label{eq_density_time_deriv} 
\end{eqnarray}

Equation (\ref{eq_einstein_nodiagonal_component}) can also be combined 
with the
time-time component of the Einstein equation to give
in three more steps:
\begin{eqnarray}
\frac{1}{2}G^{tt}=F+2\overline{F} & = & \frac{1}{2r^{2}}\left( 
1+u^{2}-\Gamma ^{2}\right) -\frac{\Gamma }{r}\left[ \frac{\Gamma 
'}{r'}-\frac{u}{\alpha \Gamma }\left( 
\frac{\dot{r}'}{r'}-\frac{\dot{\Gamma }}{\Gamma }\right) \right] ,
\nonumber \\
\frac{1}{2}r^{2}r'G^{tt} & = & \frac{r'}{2}\left( 1+u^{2}-\Gamma 
^{2}\right) +rr'\left[ \frac{\Gamma \Gamma 
'}{r'}-\frac{uu'}{r'}-\frac{4\pi ruq}{\Gamma }\right] ,
\nonumber \\
\left[ \frac{r}{2}\left( 1+u^{2}-\Gamma ^{2}\right) \right] ^{\prime 
} & = & 4\pi r^{2}r'\rho \left( 1+e\right) +\frac{4\pi 
r^{2}r'uq}{\Gamma }.\label{eq_tt_component}
\end{eqnarray}
It is now convenient to define a ``gravitational mass'' \( 
m=r/2\left( 1+u^{2}-\Gamma ^{2}\right)  \)
that leads, with the help of relation (\ref{eq_radius_space_deriv})
and (\ref{eq_tt_component}), to the concise
expressions
\begin{eqnarray}
\Gamma  & = & \sqrt{1+u^{2}-\frac{2m}{r}}\label{eq_gamma_definition} 
\\
m' & = & \Gamma \left( 1+e\right) +\frac{uq}{\rho 
}.\label{eq_mass_space_deriv} 
\end{eqnarray}
The physical interpretation of \( m(a,t) \) as the total energy 
of the
fluid inside coordinate \( a \) becomes clear if one looks at an 
evolution
equation for the mass \( m \). In order to derive such an equation,
we take the time
derivative of Eq. (\ref{eq_gamma_definition}). We already have the 
time derivative
of \( \Gamma  \) in Eq. (\ref{eq_gamma_time_deriv}) and the time 
derivative
of \( r \) in the definition of the velocity, but we still need to 
calculate
the time derivative of the velocity before being able to isolate the 
time derivative
of \( m \). An evolution equation for the velocity can be deduced 
from the
space-space component of the Einstein equations: 
\begin{eqnarray}
\frac{1}{2}G^{aa}=-F-2\overline{E} & = & 
-\frac{1}{r^{2}}\frac{m}{r}-\frac{1}{r}\left( \frac{\dot{u}}{\alpha 
}-\Gamma ^{2}\frac{\Phi '}{r'}\right) =4\pi \left( p+Q\right) 
,\nonumber \\
\frac{\dot{u}}{\alpha } & = & \Gamma ^{2}\frac{\Phi 
'}{r'}-\frac{m}{r^{2}}-4\pi r\left( p+Q\right) 
.\label{eq_velocity_time_derivative} 
\end{eqnarray}
Therefore, the evolution of the gravitational mass is
\begin{equation}
\frac{\partial m}{\alpha \partial t} = -4\pi r^{2}\left( u\left( p+Q\right) 
+\Gamma q\right) .\label{eq_mass_time_deriv} 
\end{equation}
The change of the total energy \( m \) within a sphere is given by 
surface work against the pressure and the in-out flow of energy by
transport processes.

Analogous to the derivation of the density evolution equation 
(\ref{eq_density_time_deriv})
it is possible to take the time derivative of Eq. 
(\ref{eq_mass_space_deriv})
and use Eq. (\ref{eq_mass_time_deriv}) together with one of the 
angular components
of the Einstein equations to derive an evolution equation 
for the specific energy. However, we choose to 
derive these
equations more simply from the vanishing four-divergence of the 
stress-energy tensor.
First, we list out the nonvanishing connection coefficients from 
Eqs. (\ref{eq_connection_forms})
and the definition \( \omega _{\mu \nu }=\Gamma _{\mu \nu \alpha 
}\omega ^{\alpha } \) that are used in the following computation
of the four-divergence:
\begin{eqnarray*}
\Gamma \frac{\Phi '}{r'} & = & \Gamma _{\: at}^{t}=\Gamma _{\: 
tt}^{a}\\
\rho \frac{\partial}{\alpha \partial t}\left( \frac{1}{\rho }\right) -\frac{2u}{r} & 
= & \Gamma _{\: aa}^{t}=\Gamma _{\: ta}^{a}\\
\frac{u}{r} & = & \Gamma _{\: \vartheta \vartheta }^{t}=\Gamma _{\: 
t\vartheta }^{\vartheta }=\Gamma _{\: \varphi \varphi }^{t}=\Gamma 
_{\: t\varphi }^{\varphi }\\
\frac{\Gamma }{r} & = & -\Gamma _{\: \vartheta \vartheta }^{a}=\Gamma 
_{\: a\vartheta }^{\vartheta }=-\Gamma _{\: \varphi \varphi 
}^{a}=\Gamma _{\: a\varphi }^{\varphi }\\
\frac{1}{r}\frac{\cos \vartheta }{\sin \vartheta } & = & -\Gamma _{\: 
\varphi \varphi }^{\vartheta }=\Gamma _{\: \vartheta \varphi 
}^{\varphi }.
\end{eqnarray*}
Note that we have to respect \( \Gamma _{\: \mu \nu }^{\alpha }\neq 
\Gamma _{\: \nu \mu }^{\alpha } \)
in our noncoordinate basis (\ref{eq_orthonormal_basis}).
With this in mind, one easily
obtains an evolution equation for the specific energy from the time 
component of
the four-divergence:
\begin{eqnarray}
0 & = & \frac{1}{\rho }T_{\: \: ;\nu }^{t\nu }\nonumber \\
 & = & \frac{\partial e}{\alpha \partial t}+\left( p+Q\right) \frac{\partial}{\alpha 
\partial t}\left( \frac{1}{\rho }\right) +\frac{1}{\alpha ^{2}}\left( 4\pi 
r^{2}q\alpha ^{2}\right) ^{\prime }-\frac{3u}{r}\frac{Q}{\rho 
}.\label{eq_internal_energy_time_deriv} 
\end{eqnarray}
The last equation - usually used to determine \( \Phi \) and the lapse 
function \( \alpha = e^{\Phi} \) -
is then derived from the space component of the four-divergence:
\begin{eqnarray}
0 & = & \frac{r'}{\Gamma }\frac{1}{\rho }T_{\: \: ;\nu }^{a\nu 
}\nonumber \\
 & = & \frac{\left( p+Q\right) ^{\prime }}{\rho }+\left( 
1+e+\frac{p+Q}{\rho }\right) \Phi '+\frac{\partial}{\alpha \partial t}\left( 
\frac{1}{4\pi r^{2}\rho }\frac{q}{\rho }\right) 
+\frac{3r'}{r}\frac{Q}{\rho }.\label{eq_phi_space_deriv} 
\end{eqnarray}
This set of equations (without the artificial viscosity) was, with 
some approximations, derived by Misner and Sharp
\cite{Misner_Sharp_64,Misner_Sharp_65}.

\end{document}